\documentclass[conference]{IEEEtran}
\usepackage{amsmath,amssymb,amsfonts,amsthm}
\usepackage{algorithm,algorithmic}
\usepackage{graphicx,pgf,tikz}
\usetikzlibrary{arrows,automata}
\IEEEoverridecommandlockouts
\theoremstyle{definition}
\newtheorem{theorem}{Theorem}
\newtheorem{corollary}{Corollary}

\allowdisplaybreaks

\begin{document}
\title{Delay-Optimal Buffer-Aware Probabilistic Scheduling with Adaptive Transmission}
\author{
\IEEEauthorblockN{Xiang Chen, \IEEEmembership{Student Member, IEEE}, Wei Chen, \IEEEmembership{Senior Member, IEEE}}
\IEEEauthorblockA{
Tsinghua National Laboratory for Information Science and Technology (TNList)\\
Department of Electronic Engineering, Tsinghua University, Beijing, 100084, CHINA\\
Email: chen-xiang12@mails.tsinghua.edu.cn, wchen@tsinghua.edu.cn}}
\maketitle

\begin{abstract}
Cross-layer scheduling is a promising way to improve Quality of Service (QoS) given a power constraint. In this paper, we investigate the system with random data arrival and adaptive transmission. Probabilistic scheduling strategies aware of the buffer state are applied to generalize conventional deterministic scheduling. Based on this, the average delay and power consumption are analysed by Markov reward process. The optimal delay-power tradeoff curve is the Pareto frontier of the feasible delay-power region. It is proved that the optimal delay-power tradeoff is piecewise-linear, whose vertices are obtained by deterministic strategies. Moreover, the corresponding strategies of the optimal tradeoff curve are threshold-based, hence can be obtained by a proposed effective algorithm. On the other hand, we formulate a linear programming to minimize the average delay given a fixed power constraint. By varying the power constraint, the optimal delay-power tradeoff curve can also be obtained. It is demonstrated that the algorithm result and the optimization result match each other, and are further validated by Monte-Carlo simulation.
\end{abstract}

\section{Introduction}
Real-time services such as instant messenger (IM), social network service (SNS) and streaming media are experiencing a period of rapid growth, therefore are specially concerned by next generation networks. Services of this sort all have high requirements for manipulated delay. However, due to the randomness of the workload and the architecture of traditional packet-switched communication network, delay is not easy to analyse or control. It is widely acknowledged that there is a fundamental tradeoff between the average queueing delay and the power consumption. Therefore to achieve better QoS, cross-layer optimization with traffic and energy management is well studied and accepted.

One of the major methods of achieving the optimal delay is dynamic programming. To our best knowledge, \cite{collins1999transmission} is one of the earliest work to study cross-layer scheduling model considering a time-varying channel and average delay constraints. Berry and Gallager investigated adapting the user's transmission rate and power over a fading channel in \cite{berry2002communication}. The optimal power-delay tradeoff was achieved using Markov decision process formulation and dynamic programming. Based on this, the existence of stationary optimal scheduling as well as the lower and upper bounds of the optimal policy was discovered in \cite{goyal2003power}. Detailed power/rate adaptation was further studied in \cite{bettesh2006optimal}, where analytic analysis for asymptotically large buffer size was conducted. In \cite{ata2005dynamic}, Ata modelled the problem as a continuous-time Markov chain under the assumption of fixed channel, Poisson arrival and exponentially distributed packet size. Instead of considering average delay, hard deadline as well as static channel assumption was considered in \cite{uysal2002energy}, where lazy scheduling was proposed to minimize the total energy consumption. A cumulative curves methodology was investigated in \cite{zafer2009calculus} so that energy-efficient transmission with deadline constraint and continuous-time optimization can be achieved. In \cite{lee2009energy}, closed-form optimal policy for two timeslots and numerical optimal policy for more timeslots were obtained, which were found to be a linear combination of a delay-associated term and a channel-aware term. ’K-block’ delay constraint with causal feedback of the channel state was imposed in \cite{negi2002delay}.

In our previous work \cite{chen2007optimal}, we investigated the situation where data packets arrive randomly and the channel is time-varying. We proposed a cross-layer method to achieve closed-form delay-power tradeoff and optimal scheduling strategy by a fixed-modulation scheme. Furthermore, in order to apply our formulation to various problems, the long-coherent-time and short-coherent-time cases were studied respectively in \cite{chen2014joint} and \cite{chen2015joint}. All the above work is based on simple communication protocols with fixed modulation and coding scheme so that the computation and energy cost can be saved. However, adaptive transmission rate is widely supported for many application scenarios. Therefore we advance a new approach to analysing the system with adaptive transmission in this paper.

The remainder of this paper is organized as follows. The system model is established in Section II. We formulate the problem as a Markov reward process in Section III, based on which the average power consumption and the average delay are analysed and expressed by steady-state probability distribution. In Section IV we investigate some properties of the scheduling strategies. It is discovered that the feasible delay-power region is a polyhedron, whose vertices are obtained by deterministic scheduling. The optimal delay-power tradeoff curve, as the Pareto frontier of the polyhedron, is piecewise linear. Moreover, the optimal scheduling is threshold-based, thus an algorithm is proposed to obtain the optimal tradeoff curve. In Section V, we formulate a linear programming to minimize the average delay given an average power constraint. By varying the power constraint, the optimal tradeoff curve can also be acquired. It is demonstrated that the algorithm, the optimization and the simulation results completely match each other in Section VI. Section VII concludes the paper.

\section{System Model}
\begin{figure}[t]
\centering
\includegraphics[width=1\columnwidth]{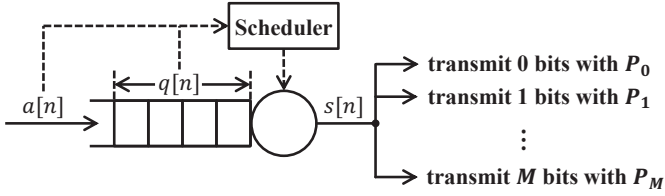}
\caption{System Model}
\label{fig_model}
\end{figure}

Consider the system model shown in \figurename \ref{fig_model}. We assume that at the beginning of each timeslot, data arrive as a Bernoulli Process with parameter $\alpha$. Each incoming data packet contains $A$ bits. Denote $a[n]=1$ or $0$ as whether or not there are data arriving in timeslot $n$, hence
\begin{align}
\begin{cases}
\text{Pr}\{a[n]=1\}=\alpha,\\
\text{Pr}\{a[n]=0\}=1-\alpha.
\end{cases}
\end{align}

Denote $s[n]$ as the number of data bits transmitted in timeslot $n$. Because of the constraints of the transmitter, in each timeslot at most $M$ bits can be transmitted. In order to guarantee the stability of the system, we set $M \ge A$. Assume transmitting $m$ bits will cost power $P_m$. Transmitting $0$ bit will cost no power, hence $P_0=0$. From perspective of the physical layer, to transmit more bits without increasing the error rate, we should use a larger constellation diagram, therefore more power has to be consumed for every bit in average, i.e. $\frac{P_{m_1}}{m_1}<\frac{P_{m_2}}{m_2}$ if $m_1<m_2$. More details can be found in \cite{uysal2002energy}, where information-theoretic explanations and a few examples are provided to support this point.

To temporarily store the backlog data, a buffer is introduced. Assume the buffer can restore at most $Q$ bits. Denote $q[n]$ as the number of bits in the buffer at the beginning of timeslot $n$. Therefore in timeslot $n$, we can decide how many bits to be transmitted based on the buffer state $q[n]$ and the data arrival $a[n]$. To extend from traditional deterministic scheduling methods so that a better delay-power tradeoff curve can be achieved, we investigate probabilistic scheduling policies. In other words, given $q[n]$ and $a[n]$, the value of $s[n]$ is determined by a probability distribution. In each timeslot, the transmission is scheduled after the data arrival, thus when deciding how many bits should be transmitted, there are $(q[n]+Aa[n])$ bits data in total. Denote $t[n]=q[n]+Aa[n]$ and $K=Q+A$. For transmission, there is no need to distinguish the newly arrived data from the backlog data, therefore we denote $f_{k,m}$ as the probability to transmit $m$ bits when $t[n]=k$, i.e.
\begin{align}
f_{k,m}=\text{Pr}\{s[n]=m|t[n]=k\}.
\end{align}
Immediately we have
\begin{align}
\sum_{m=0}^{M}f_{k,m}=1 \qquad \forall k=0,\cdots,K.
\label{sum=1}
\end{align}
Denote $\boldsymbol{F}$ as a $(K+1)\times(M+1)$ matrix whose element in the $(k+1)$th row and the $(m+1)$th column is $f_{k,m}$. We guarantee that the transmission strategy will avoid overflow or underflow, which means $f_{k,m}=0$ if $k-m<0$ or $k-m>Q$. Therefore the following equation should hold
\begin{align}
&q[n+1]=q[n]+Aa[n]-s[n],\\
&t[n+1]=t[n]-s[n]+Aa[n+1].\label{t}
\end{align}

\begin{figure*}[!t]
\centering
\begin{tikzpicture}[->, >=stealth', very thick, every node/.style={fill=white, font=\normalsize}]
\node[state] (0) at (0,0) {$0$};
\node[state] (1) at (2.4,0) {$1$};
\node[state] (2) at (4.8,0) {$2$};
\node[state] (3) at (7.2,0) {$3$};
\node[state] (4) at (9.6,0) {$4$};
\node[state] (5) at (12,0) {$5$};
\node[state] (6) at (14.4,0) {$6$};
\node[state] (7) at (16.8,0) {$7$};

\path
(1) edge [bend right=15] node {$\lambda_{1,0}$} (0)
(2) edge [bend right=15] node {$\lambda_{2,1}$} (1)
(3) edge [bend right=15] node {$\lambda_{3,2}$} (2)
(4) edge [bend right=15] node {$\lambda_{4,3}$} (3)
(5) edge [bend right=15] node {$\lambda_{5,4}$} (4)
(6) edge [bend right=15] node {$\lambda_{6,5}$} (5)
(7) edge [bend right=15] node {$\lambda_{7,6}$} (6)

(2) edge [bend right=45] node {$\lambda_{2,0}$} (0)
(3) edge [bend right=45] node {$\lambda_{3,1}$} (1)
(4) edge [bend right=45] node {$\lambda_{4,2}$} (2)
(5) edge [bend right=45] node {$\lambda_{5,3}$} (3)
(6) edge [bend right=45] node {$\lambda_{6,4}$} (4)
(7) edge [bend right=45] node {$\lambda_{7,5}$} (5)

(3) edge [bend right=75] node {$\lambda_{3,0}$} (0)
(4) edge [bend right=75] node {$\lambda_{4,1}$} (1)
(5) edge [bend right=75] node {$\lambda_{5,2}$} (2)
(6) edge [bend right=75] node {$\lambda_{6,3}$} (3)
(7) edge [bend right=75] node {$\lambda_{7,4}$} (4)

(0) edge [bend right=15] node {$\lambda_{0,1}$} (1)
(1) edge [bend right=15] node {$\lambda_{1,2}$} (2)
(2) edge [bend right=15] node {$\lambda_{2,3}$} (3)
(3) edge [bend right=15] node {$\lambda_{3,4}$} (4)
(4) edge [bend right=15] node {$\lambda_{4,5}$} (5)
(5) edge [bend right=15] node {$\lambda_{5,6}$} (6)
(6) edge [bend right=15] node {$\lambda_{6,7}$} (7)

(0) edge [bend right=45] node {$\lambda_{0,2}$} (2)
(1) edge [bend right=45] node {$\lambda_{1,3}$} (3)
(2) edge [bend right=45] node {$\lambda_{2,4}$} (4)
(3) edge [bend right=45] node {$\lambda_{3,5}$} (5)
(4) edge [bend right=45] node {$\lambda_{4,6}$} (6)
(5) edge [bend right=45] node {$\lambda_{5,7}$} (7)
;
\end{tikzpicture}
\caption{Markov Chain of $t[n]$ ($K=7$, $A=2$, $M=3$)}
\label{fig_markov}
\end{figure*}
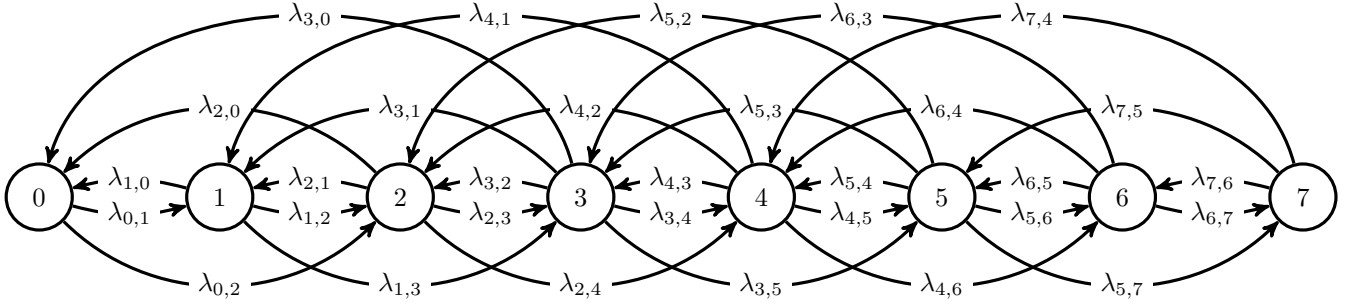

\section{Markov Reward Process}
As described in the above section, the scheduler will decide how many bits to be transmitted based on the information of the queue length and the data arrival state. Since $a[n]$ is $i.i.d.$, $s[n]$ is determined by $t[n]$, based on (\ref{t}) we have
\begin{align}
&\text{Pr}\{t[n+1]|t[n],\cdots,t[1]\}=\text{Pr}\{t[n+1]|t[n]\}.
\end{align}
Hence the stochastic process $\{t[n]\}$ is Markovian. Therefore we can formulate the problem as a Markov Reward Process (MRP), where the average delay and power are the rewards of the Markov process. Denote $\pi_k$ as the steady-state probability for $t[n]=k$. Denote $\lambda_{i,j}$ as the transition probability from state $i$ to state $j$. The transition diagram is shown in \figurename \ref{fig_markov}, where $\lambda_{i,i}$ for $i=0,\cdots,K$ are omitted to keep the diagram legible. The transition from state $i$ to state $j$ is made up of at most two cases. One is transmitting $(i-j)$ bits and then there are no data arriving. The other is transmitting $(i-j+A)$ bits and then there are data arriving. Therefore we have
\begin{align}
&\lambda_{i,j}=\nonumber\\
&\begin{cases}
(1-\alpha)f_{i,i-j} & M-A < i-j \le M\\
(1-\alpha)f_{i,i-j} & 0 \le i-j \le M-A , j < A\\
\begin{array}{@{}l}
(1-\alpha)f_{i,i-j}\\
+\alpha f_{i,i-j+A}
\end{array}
 & 0 \le i-j \le M-A , A \le j \le K-A\\
\alpha f_{i,i-j+A} & 0 \le i-j \le M-A , j > K-A\\
\alpha f_{i,i-j+A} & -A \le i-j < 0 , j \ge A\\
0 & \text{else}
\end{cases}.
\end{align}

Denote $\boldsymbol{\pi}=[\pi_0,\cdots,\pi_K]^T$. Denote $\boldsymbol{\Lambda}_{\boldsymbol{F}}$ as a $(K+1)\times(K+1)$ matrix whose element in the $(i+1)$th column and the $(j+1)$th row is $\lambda_{i,j}$. It is determined by $\boldsymbol{F}$. Denote $\boldsymbol{I}$ as the identity matrix, $\boldsymbol{1}=[1,\cdots,1]^T$, and $\boldsymbol{0}=[0,\cdots,0]^T$. We won't specify their size if there is no ambiguity. Denote $\boldsymbol{G}_{\boldsymbol{F}}=\boldsymbol{\Lambda}_{\boldsymbol{F}}-\boldsymbol{I}$. Therefore we have $\boldsymbol{G}_{\boldsymbol{F}}\boldsymbol{\pi}=\boldsymbol{0}$ and $\boldsymbol{1}^T\boldsymbol{\pi}=1$. Since the sum of all rows in $\boldsymbol{G}_{\boldsymbol{F}}$ is $\boldsymbol{0}^T$, we set
$\boldsymbol{H}_{\boldsymbol{F}}=\left[
\begin{array}{c}
\boldsymbol{1}^T\\
\boldsymbol{G}_{\boldsymbol{F}}(0:(K-1),:)
\end{array}
\right]
$ and
$\boldsymbol{c}=\left[
\begin{array}{c}
1\\\boldsymbol{0}
\end{array}
\right]$. Hence
\begin{align}
\boldsymbol{\pi}=\boldsymbol{H}_{\boldsymbol{F}}^{-1}\boldsymbol{c}.
\label{pi-H}
\end{align}
In other words, $\boldsymbol{H}_{\boldsymbol{F}}$ can be used to represent a certain transmission strategy. It will determine the value of $\boldsymbol{\pi}$.

We can express the average delay and the average power consumption using the steady-state probability distribution. For state $k$, transmitting $m$ bits will cost $P_m$ with probability $f_{k,m}$. Denote $\boldsymbol{p}_{\boldsymbol{F}}=[\sum_{m=0}^M P_m f_{0,m},\cdots,\sum_{m=0}^M P_m f_{K,m}]^T$ as a function of $\boldsymbol{F}$, thus the average power consumption
\begin{align}
P=\sum_{k=0}^{K} \pi_k \sum_{m=0}^M P_m f_{k,m}=\boldsymbol{p}_{\boldsymbol{F}}^T\boldsymbol{\pi}.
\label{PwithPi}
\end{align}
Similarly, we can obtain that the average queue length is
\begin{align}
\text{Ex}\{q[n]\}=\text{Ex}\{t[n]-a[n]\}=\sum_{k=0}^{K} k \pi_k-\alpha A.
\end{align}
According to Little's Law, the average delay is the quotient of the average queue length divided by the average arrival rate. Denote $\boldsymbol{d}=[0,1,\cdots,K]^T$, we have
\begin{align}
D=\frac{1}{\alpha A}\left(\sum_{k=0}^{K} k \pi_k-\alpha A\right)=\frac{1}{\alpha A}\boldsymbol{d}^T\boldsymbol{\pi}-1.
\label{DwithPi}
\end{align}

\section{Properties of Scheduling Strategies}
From above it can be seen that $\boldsymbol{p}_{\boldsymbol{F}}$ and $\boldsymbol{H}_{\boldsymbol{F}}$ are determined by $\boldsymbol{F}$. Moreover, $\boldsymbol{\pi}$ is determined by $\boldsymbol{H}_{\boldsymbol{F}}$. Therefore a certain scheduling strategy $\boldsymbol{F}$ will determine a delay-power pair $(P,D)$. Denote $(P_{\boldsymbol{F}},D_{\boldsymbol{F}})$ as the delay-power pair generated by strategy $\boldsymbol{F}$. The mapping from $\boldsymbol{F}$ to $(P_{\boldsymbol{F}},D_{\boldsymbol{F}})$ has the following property.

\begin{theorem}
$\boldsymbol{F}$ and $\boldsymbol{F}'$ are two scheduling strategies that are different only when $t[n]=k$, i.e. the two matrices are different only in the $(k+1)$th row. Denote $\boldsymbol{F}''=(1-\epsilon)\boldsymbol{F}+\epsilon\boldsymbol{F}'$ where $0\le \epsilon\le 1$. Then

1) There exists a certain $0\le \epsilon'\le 1$ so that $P_{\boldsymbol{F}''}=(1-\epsilon')P_{\boldsymbol{F}}+\epsilon' P_{\boldsymbol{F}'}$ and $D_{\boldsymbol{F}''}=(1-\epsilon')D_{\boldsymbol{F}}+\epsilon' D_{\boldsymbol{F}'}$. Moreover, $\epsilon'$ is a continuous nondecreasing function of $\epsilon$.

2) When $\epsilon$ changes from 0 to 1, point $(P_{\boldsymbol{F}''},D_{\boldsymbol{F}''})$ moves on the line segment from $(P_{\boldsymbol{F}},D_{\boldsymbol{F}})$ to $(P_{\boldsymbol{F}'},D_{\boldsymbol{F}'})$.
\label{theorem_linearcombination}
\end{theorem}

\begin{IEEEproof}
We will prove the two conclusions one by one.

1) From the definition of $\boldsymbol{H}_{\boldsymbol{F}}$ and $\boldsymbol{p}_{\boldsymbol{F}}$, we can see that if $\boldsymbol{F}''=(1-\epsilon)\boldsymbol{F}+\epsilon\boldsymbol{F}'$, then $\boldsymbol{H}_{\boldsymbol{F}''}=(1-\epsilon)\boldsymbol{H}_{\boldsymbol{F}}+\epsilon\boldsymbol{H}_{\boldsymbol{F}'}$ and $\boldsymbol{p}_{\boldsymbol{F}''}=(1-\epsilon)\boldsymbol{p}_{\boldsymbol{F}}+\epsilon\boldsymbol{p}_{\boldsymbol{F}'}$. Denote $\Delta\boldsymbol{H}=\boldsymbol{H}_{\boldsymbol{F}'}-\boldsymbol{H}_{\boldsymbol{F}}$ and $\Delta\boldsymbol{p}=\boldsymbol{p}_{\boldsymbol{F}'}-\boldsymbol{p}_{\boldsymbol{F}}$. Since $\boldsymbol{F}$ and $\boldsymbol{F}'$ are different only in the $(k+1)$th row, it can be derived that $\Delta \boldsymbol{H}$ has nonzero element only in the $(k+1)$th column, and the $(k+1)$th element of $\Delta \boldsymbol{p}$ is its only nonzero element. Therefore $\Delta \boldsymbol{H}$ can be denoted as $\left[\boldsymbol{0},\cdots,\boldsymbol{\delta}_k,\cdots,\boldsymbol{0}\right]$, where $\boldsymbol{\delta}_k$ is its $(k+1)$th column. $\Delta \boldsymbol{p}$ can be denoted as $\left[0,\cdots,\zeta_k,\cdots,0\right]^T$, where $\zeta_k$ is its $(k+1)$th element. Also, we denote $\boldsymbol{H}_{\boldsymbol{F}}^{-1}=\left[
\begin{array}{c}
\boldsymbol{h}_0^T\\
\boldsymbol{h}_1^T\\
\vdots\\
\boldsymbol{h}_K^T
\end{array}
\right]$. Hence
\begin{align}
(\boldsymbol{H}_{\boldsymbol{F}}^{-1}\Delta\boldsymbol{H})\boldsymbol{H}_{\boldsymbol{F}}^{-1}=\left[
\begin{array}{c}
(\boldsymbol{h}_0^T\boldsymbol{\delta}_k)\boldsymbol{h}_k^T\\
(\boldsymbol{h}_1^T\boldsymbol{\delta}_k)\boldsymbol{h}_k^T\\
\vdots\\
(\boldsymbol{h}_K^T\boldsymbol{\delta}_k)\boldsymbol{h}_k^T
\end{array}
\right].
\end{align}

By mathematical induction, we can have that for $i\ge 1$,
\begin{align}
&(\boldsymbol{H}_{\boldsymbol{F}}^{-1}\Delta\boldsymbol{H})^{i}\boldsymbol{H}_{\boldsymbol{F}}^{-1}\nonumber\\
=&\left[
\begin{array}{c}
(\boldsymbol{h}_0^T\boldsymbol{\delta}_k)(\boldsymbol{h}_k^T\boldsymbol{\delta}_k)^{i-1}\boldsymbol{h}_k^T\\
(\boldsymbol{h}_1^T\boldsymbol{\delta}_k)(\boldsymbol{h}_k^T\boldsymbol{\delta}_k)^{i-1}\boldsymbol{h}_k^T\\
\vdots\\
(\boldsymbol{h}_K^T\boldsymbol{\delta}_k)(\boldsymbol{h}_k^T\boldsymbol{\delta}_k)^{i-1}\boldsymbol{h}_k^T
\end{array}
\right]\\
=&(\boldsymbol{h}_k^T\boldsymbol{\delta}_k)^{i-1}(\boldsymbol{H}_{\boldsymbol{F}}^{-1}\Delta\boldsymbol{H})\boldsymbol{H}_{\boldsymbol{F}}^{-1}
\end{align}
and
\begin{align}
&\Delta\boldsymbol{p}^T\boldsymbol{H}_{\boldsymbol{F}}^{-1}(\boldsymbol{H}_{\boldsymbol{F}}^{-1}\Delta\boldsymbol{H})^{i-1}\nonumber\\
=&\zeta_k(\boldsymbol{h}_k^T\boldsymbol{\delta}_k)^{i-1}\boldsymbol{h}_k^T.
\end{align}

Therefore the expansion
\begin{align}
&(\boldsymbol{H}_{\boldsymbol{F}}+\epsilon\Delta\boldsymbol{H})^{-1}\nonumber\\
=&\sum_{i=0}^{+\infty} (-\epsilon)^{i}(\boldsymbol{H}_{\boldsymbol{F}}^{-1}\Delta\boldsymbol{H})^{i}\boldsymbol{H}_{\boldsymbol{F}}^{-1}\\
=&\boldsymbol{H}_{\boldsymbol{F}}^{-1}
+\sum_{i=1}^{+\infty}(-\epsilon)^i(\boldsymbol{h}_k^T\boldsymbol{\delta}_k)^{i-1}(\boldsymbol{H}_{\boldsymbol{F}}^{-1}\Delta\boldsymbol{H})\boldsymbol{H}_{\boldsymbol{F}}^{-1}.
\end{align}

From (\ref{pi-H}), (\ref{PwithPi}) and (\ref{DwithPi}), we have $P_{\boldsymbol{F}}=\boldsymbol{p}_{\boldsymbol{F}}^T\boldsymbol{H}_{\boldsymbol{F}}^{-1}\boldsymbol{c}$ and $D_{\boldsymbol{F}}=\frac{1}{\alpha A}\boldsymbol{d}^T\boldsymbol{H}_{\boldsymbol{F}}^{-1}\boldsymbol{c}-1$. Therefore
\begin{align}
&\frac{P_{\boldsymbol{F}''}-P_{\boldsymbol{F}}}{P_{\boldsymbol{F}'}-P_{\boldsymbol{F}}}\nonumber\\
=&\frac{(\boldsymbol{p}_{\boldsymbol{F}}+\epsilon\Delta\boldsymbol{p})^T(\boldsymbol{H}_{\boldsymbol{F}}+\epsilon\Delta\boldsymbol{H})^{-1}\boldsymbol{c}-\boldsymbol{p}_{\boldsymbol{F}}^T\boldsymbol{H}_{\boldsymbol{F}}^{-1}\boldsymbol{c}}{(\boldsymbol{p}_{\boldsymbol{F}}+\Delta\boldsymbol{p})^T(\boldsymbol{H}_{\boldsymbol{F}}+\Delta\boldsymbol{H})^{-1}\boldsymbol{c}-\boldsymbol{p}_{\boldsymbol{F}}^T\boldsymbol{H}_{\boldsymbol{F}}^{-1}\boldsymbol{c}}\\
=&\frac{
\begin{array}{c}
\boldsymbol{p}_{\boldsymbol{F}}^T\left[(\boldsymbol{H}_{\boldsymbol{F}}+\epsilon\Delta\boldsymbol{H})^{-1}-\boldsymbol{H}_{\boldsymbol{F}}^{-1}\right]\boldsymbol{c}\\
+\epsilon\Delta\boldsymbol{p}^T(\boldsymbol{H}_{\boldsymbol{F}}+\epsilon\Delta\boldsymbol{H})^{-1}\boldsymbol{c}
\end{array}
}
{
\begin{array}{c}
\boldsymbol{p}_{\boldsymbol{F}}^T\left[(\boldsymbol{H}_{\boldsymbol{F}}+\Delta\boldsymbol{H})^{-1}-\boldsymbol{H}_{\boldsymbol{F}}^{-1}\right]\boldsymbol{c}\\
+\Delta\boldsymbol{p}^T(\boldsymbol{H}_{\boldsymbol{F}}+\Delta\boldsymbol{H})^{-1}\boldsymbol{c}
\end{array}
}\\
=&\frac{
\begin{array}{c}
\boldsymbol{p}_{\boldsymbol{F}}^T\left[\sum_{i=1}^{+\infty}(-\epsilon)^i(\boldsymbol{h}_k^T\boldsymbol{\delta}_k)^{i-1}(\boldsymbol{H}_{\boldsymbol{F}}^{-1}\Delta\boldsymbol{H})\boldsymbol{H}_{\boldsymbol{F}}^{-1}\right]\boldsymbol{c}\\
-\Delta\boldsymbol{p}^T\left[\sum_{i=1}^{+\infty} (-\epsilon)^{i}(\boldsymbol{H}_{\boldsymbol{F}}^{-1}\Delta\boldsymbol{H})^{i-1}\boldsymbol{H}_{\boldsymbol{F}}^{-1}\right]\boldsymbol{c}
\end{array}
}
{
\begin{array}{c}
\boldsymbol{p}_{\boldsymbol{F}}^T\left[\sum_{i=1}^{+\infty}(-1)^i(\boldsymbol{h}_k^T\boldsymbol{\delta}_k)^{i-1}(\boldsymbol{H}_{\boldsymbol{F}}^{-1}\Delta\boldsymbol{H})\boldsymbol{H}_{\boldsymbol{F}}^{-1}\right]\boldsymbol{c}\\
-\Delta\boldsymbol{p}^T\left[\sum_{i=1}^{+\infty} (-1)^{i}(\boldsymbol{H}_{\boldsymbol{F}}^{-1}\Delta\boldsymbol{H})^{i-1}\boldsymbol{H}_{\boldsymbol{F}}^{-1}\right]\boldsymbol{c}
\end{array}
}\\
=&\frac{
\begin{array}{c}
\sum_{i=1}^{+\infty}(-\epsilon)^i(\boldsymbol{h}_k^T\boldsymbol{\delta}_k)^{i-1}\boldsymbol{p}_{\boldsymbol{F}}^T(\boldsymbol{H}_{\boldsymbol{F}}^{-1}\Delta\boldsymbol{H})\boldsymbol{H}_{\boldsymbol{F}}^{-1}\boldsymbol{c}\\
-\sum_{i=1}^{+\infty} (-\epsilon)^{i}\zeta_k(\boldsymbol{h}_k^T\boldsymbol{\delta}_k)^{i-1}\boldsymbol{h}_k^T\boldsymbol{c}
\end{array}
}
{
\begin{array}{c}
\sum_{i=1}^{+\infty}(-1)^i(\boldsymbol{h}_k^T\boldsymbol{\delta}_k)^{i-1}\boldsymbol{p}_{\boldsymbol{F}}^T(\boldsymbol{H}_{\boldsymbol{F}}^{-1}\Delta\boldsymbol{H})\boldsymbol{H}_{\boldsymbol{F}}^{-1}\boldsymbol{c}\\
-\sum_{i=1}^{+\infty} (-1)^{i}\zeta_k(\boldsymbol{h}_k^T\boldsymbol{\delta}_k)^{i-1}\boldsymbol{h}_k^T\boldsymbol{c}
\end{array}
}\\
=&\frac{\sum_{i=1}^{+\infty}(-\epsilon)^i(\boldsymbol{h}_k^T\boldsymbol{\delta}_k)^{i-1}}
{\sum_{i=1}^{+\infty}(-1)^i(\boldsymbol{h}_k^T\boldsymbol{\delta}_k)^{i-1}}\\
=&\frac{\epsilon+\epsilon\boldsymbol{h}_k^T\boldsymbol{\delta}_k}{1+\epsilon\boldsymbol{h}_k^T\boldsymbol{\delta}_k}
\end{align}
and
\begin{align}
&\frac{D_{\boldsymbol{F}''}-D_{\boldsymbol{F}}}{D_{\boldsymbol{F}'}-D_{\boldsymbol{F}}}\nonumber\\
=&\frac{\boldsymbol{d}^T(\boldsymbol{H}_{\boldsymbol{F}}+\epsilon\Delta\boldsymbol{H})^{-1}\boldsymbol{c}-\boldsymbol{d}^T\boldsymbol{H}_{\boldsymbol{F}}^{-1}\boldsymbol{c}}{\boldsymbol{d}^T(\boldsymbol{H}_{\boldsymbol{F}}+\Delta\boldsymbol{H})^{-1}\boldsymbol{c}-\boldsymbol{d}^T\boldsymbol{H}_{\boldsymbol{F}}^{-1}\boldsymbol{c}}\\
=&\frac{\boldsymbol{d}^T(\sum_{i=1}^{+\infty}(-\epsilon)^i(\boldsymbol{h}_k^T\boldsymbol{\delta}_k)^{i-1}(\boldsymbol{H}_{\boldsymbol{F}}^{-1}\Delta\boldsymbol{H})\boldsymbol{H}_{\boldsymbol{F}}^{-1})\boldsymbol{c}}{\boldsymbol{d}^T(\sum_{i=1}^{+\infty}(-1)^i(\boldsymbol{h}_k^T\boldsymbol{\delta}_k)^{i-1}(\boldsymbol{H}_{\boldsymbol{F}}^{-1}\Delta\boldsymbol{H})\boldsymbol{H}_{\boldsymbol{F}}^{-1})\boldsymbol{c}}\\
=&\frac{\sum_{i=1}^{+\infty}(-\epsilon)^i(\boldsymbol{h}_k^T\boldsymbol{\delta}_k)^{i-1}}{\sum_{i=1}^{+\infty}(-1)^i(\boldsymbol{h}_k^T\boldsymbol{\delta}_k)^{i-1}}\\
=&\frac{\epsilon+\epsilon\boldsymbol{h}_k^T\boldsymbol{\delta}_k}{1+\epsilon\boldsymbol{h}_k^T\boldsymbol{\delta}_k}.
\end{align}
Hence $\frac{P_{\boldsymbol{F}''}-P_{\boldsymbol{F}}}{P_{\boldsymbol{F}'}-P_{\boldsymbol{F}}}=\frac{D_{\boldsymbol{F}''}-D_{\boldsymbol{F}}}{D_{\boldsymbol{F}'}-D_{\boldsymbol{F}}}=\frac{\epsilon+\epsilon\boldsymbol{h}_k^T\boldsymbol{\delta}_k}{1+\epsilon\boldsymbol{h}_k^T\boldsymbol{\delta}_k}=\epsilon'$, so that $P_{\boldsymbol{F}''}=(1-\epsilon')P_{\boldsymbol{F}}+\epsilon' P_{\boldsymbol{F}'}$ and $D_{\boldsymbol{F}''}=(1-\epsilon')D_{\boldsymbol{F}}+\epsilon' D_{\boldsymbol{F}'}$. Moreover, it can be observed that $\epsilon'=\frac{\epsilon+\epsilon\boldsymbol{h}_k^T\boldsymbol{\delta}_k}{1+\epsilon\boldsymbol{h}_k^T\boldsymbol{\delta}_k}$ is a continuous nondecreasing function.

2) From the first part, we know $\frac{P_{\boldsymbol{F}''}-P_{\boldsymbol{F}}}{P_{\boldsymbol{F}'}-P_{\boldsymbol{F}}}=\frac{D_{\boldsymbol{F}''}-D_{\boldsymbol{F}}}{D_{\boldsymbol{F}'}-D_{\boldsymbol{F}}}=\epsilon'$ and $\epsilon'$ is a continuous nondecreasing function of $\epsilon$. When $\epsilon=0$, $\epsilon'=0$. When $\epsilon=1$, $\epsilon'=1$. Therefore when $\epsilon$ changes from 0 to 1, the point $(P_{\boldsymbol{F}''},D_{\boldsymbol{F}''})$ moves on the line segment from $(P_{\boldsymbol{F}},D_{\boldsymbol{F}})$ to $(P_{\boldsymbol{F}'},D_{\boldsymbol{F}'})$. The slope of the line is
\begin{align}
&\frac{D_{\boldsymbol{F}'}-D_{\boldsymbol{F}}}{P_{\boldsymbol{F}'}-P_{\boldsymbol{F}}}\nonumber\\
=&\frac{\frac{1}{\alpha A}
\boldsymbol{d}^T(\boldsymbol{H}_{\boldsymbol{F}}+\Delta\boldsymbol{H})^{-1}\boldsymbol{c}-\frac{1}{\alpha A}\boldsymbol{d}^T\boldsymbol{H}_{\boldsymbol{F}}^{-1}\boldsymbol{c}
}{
(\boldsymbol{p}_{\boldsymbol{F}}+\Delta\boldsymbol{p})^T(\boldsymbol{H}_{\boldsymbol{F}}+\Delta\boldsymbol{H})^{-1}\boldsymbol{c}-\boldsymbol{p}_{\boldsymbol{F}}^T\boldsymbol{H}_{\boldsymbol{F}}^{-1}\boldsymbol{c}
}\\
=&\frac{\frac{1}{\alpha A}\boldsymbol{d}^T(\sum_{i=1}^{+\infty}(-1)^i(\boldsymbol{h}_k^T\boldsymbol{\delta}_k)^{i-1}(\boldsymbol{H}_{\boldsymbol{F}}^{-1}\Delta\boldsymbol{H})\boldsymbol{H}_{\boldsymbol{F}}^{-1})\boldsymbol{c}}
{
\begin{array}{c}
\sum_{i=1}^{+\infty}(-1)^i(\boldsymbol{h}_k^T\boldsymbol{\delta}_k)^{i-1}\boldsymbol{p}_{\boldsymbol{F}}^T(\boldsymbol{H}_{\boldsymbol{F}}^{-1}\Delta\boldsymbol{H})\boldsymbol{H}_{\boldsymbol{F}}^{-1}\boldsymbol{c}\\
-\sum_{i=1}^{+\infty} (-1)^{i}\zeta_k(\boldsymbol{h}_k^T\boldsymbol{\delta}_k)^{i-1}\boldsymbol{h}_k^T\boldsymbol{c}
\end{array}
}\\
=&\frac{\frac{1}{\alpha A}\boldsymbol{d}^T\boldsymbol{H}_{\boldsymbol{F}}^{-1}\Delta\boldsymbol{H}\boldsymbol{H}_{\boldsymbol{F}}^{-1}\boldsymbol{c}}
{\boldsymbol{p}_{\boldsymbol{F}}^T\boldsymbol{H}_{\boldsymbol{F}}^{-1}\Delta\boldsymbol{H}\boldsymbol{H}_{\boldsymbol{F}}^{-1}\boldsymbol{c}-\zeta_k\boldsymbol{h}_k^T\boldsymbol{c}}\\
=&\frac{\boldsymbol{d}^T\boldsymbol{H}_{\boldsymbol{F}}^{-1}\boldsymbol{\delta}_k}{\alpha A(\boldsymbol{p}_{\boldsymbol{F}}^T\boldsymbol{H}_{\boldsymbol{F}}^{-1}\boldsymbol{\delta}_k-\zeta_k)}.\label{slope}
\end{align}
\end{IEEEproof}

Theorem \ref{theorem_linearcombination} indicates that the convex combination of scheduling strategies which are different only for one $k$ will induce the convex combination of delay-power pairs. The slope of the line segment from $(P_{\boldsymbol{F}},D_{\boldsymbol{F}})$ to $(P_{\boldsymbol{F}'},D_{\boldsymbol{F}'})$ is obtained in (\ref{slope}). Furthermore, we can have the following enhanced conclusions, whose strict proofs are omitted because of space limitation.
\begin{corollary}
The set of all feasible delay-power pairs is a polyhedron. The vertices of the polyhedron are all obtained by deterministic scheduling strategies.
\label{corollary_region=polyhedron}
\end{corollary}

Denote $\mathcal{R}$ as the set of all possible delay-power pairs. Denote $\mathcal{C}$ as the convex hull of delay-power pairs generated by deterministic scheduling strategies. Corollary \ref{corollary_region=polyhedron} actually demonstrate that $\mathcal{R}=\mathcal{C}$. However, what we concern most is the optimal delay-power tradeoff curve $\mathcal{L}=\{(P,D)\in\mathcal{R}|\forall(P',D')\in\mathcal{R},\text{ either }P'\ge P\text{ or }D'\ge D\}$. In other words, $\mathcal{L}$ is the Pareto frontier of $\mathcal{R}$. Based on Corollary \ref{corollary_region=polyhedron}, immediately we can have that

\begin{corollary}
The optimal delay-power tradeoff curve is piecewise linear. The vertices of the curve are obtained by deterministic scheduling strategies.
\label{corollary_piecewiselinear}
\end{corollary}

Moreover, the scheduling strategies leading to the optimal delay-power tradeoff have the following property.

\begin{theorem}
The optimal scheduling strategy is threshold-based. That is to say, there exists $(M+1)$ thresholds $k_0 \le k_1 \le \cdots \le k_M$, one of which we denote as $k_{m^*}$, such that
\begin{align}
\begin{cases}
f_{k,m}=1 \qquad\qquad k_{m-1}<k\le k_m,m\neq m^*\\
f_{k,m}=1 \qquad\qquad k_{m^*-1}<k< k_{m^*}\\
f_{k_{m^*},m^*}+f_{k_{m^*},m^*+1}=1 \\
f_{k,m}=0 \qquad\qquad \text{else}
\end{cases}.
\end{align}
For simplicity and unification, denote $k_{-1}=-1$. More specifically, we have $k_0=0$, $k_A=k_{A+1}=\cdots=k_{M}=K$.
\label{theorem_threshold}
\end{theorem}

The proof of Theorem \ref{theorem_threshold} is also omitted because of space limitation. The theorem indicates that more data should be transmitted if the queue is longer. When $k_{m-1}<t[n]<k_{m}$, $m$ bits should be transmitted. Any optimal scheduling strategy $\boldsymbol{F}$ has at most two decimal elements $f_{k_{m^*},m^*}$ and $f_{k_{m^*},m^*+1}$, while the other elements are either 0 or 1. The strategies leading to two adjacent vertices and their line segment are only different for $t[n]=k_{m^*}$. For the vertex with larger delay and less power, $f_{k_{m^*},m^*}=1$. For the vertex with smaller delay and more power, $f_{k_{m^*},m^*+1}=1$. This confirms that the vertices are obtained by deterministic strategies.

Based on Corollary \ref{corollary_piecewiselinear} and Theorem \ref{theorem_threshold}, we know that the optimal delay-power tradeoff curve is obtained by threshold-based strategy. The vertices of the piecewise linear curve are obtained by deterministic strategy. Therefore Algorithm \ref{algo} is proposed to effectively obtain the optimal delay-power tradeoff curve. In the algorithm, we start from the bottom right vertex of the curve, and search for the next vertex on the optimal tradeoff curve one by one.

\begin{algorithm}[t]
\caption{Obtain the Optimal Delay-Power Tradeoff}
\textbf{initialize:}
\begin{enumerate}
\item Set $\boldsymbol{F}$ as a threshold-based deterministic strategy where $k_m=\min\{m,A\}$ for $0\le m\le M$
\item Calculate the corresponding average delay $d_c$ and average power consumption $p_c$
\item Denote set $\mathcal{F}_c=\{\boldsymbol{F}\}$
\end{enumerate}
\textbf{while} $\mathcal{F}_c$ is not null \textbf{do}
\begin{enumerate}
\item $\mathcal{F}_p=\mathcal{F}_c$, $d_p=d_c$, $p_p=p_c$, $\mathcal{F}_c=\emptyset$, $s=+\infty$
\item For every strategy in $\mathcal{F}_p$, we take turns to increase each one of its threshold, thus generate a new strategy $\boldsymbol{F}$ each time. For every legal scheduling strategy $\boldsymbol{F}$:\\
\textbf{if} its corresponding average delay $d$ and average power $p$ satisfies $d\ge d_p$ and $p<p_p$ and $\frac{d-d_p}{p_p-p}<s$\\
\textbf{then} $\mathcal{F}_c=\{\boldsymbol{F}\}$, $s=\frac{d-d_p}{p_p-p}$, $d_c=d$, $p_c=p$\\
\textbf{else if} its corresponding average delay $d$ and average power $p$ satisfies $d\ge d_p$ and $p<p_p$ and $\frac{d-d_p}{p_p-p}=s$\\
\textbf{then} add $\boldsymbol{F}$ into set $\mathcal{F}_c$\\
\textbf{end if}
\item Draw the line segment connecting $(p_p,d_p)$ and $(p_c,d_c)$
\end{enumerate}
\textbf{end while}
\label{algo}
\end{algorithm}

\section{Linear Programming Formulation}
In the above section, we analyse the optimal delay-power tradeoff as the Pareto frontier of the feasible delay-power region. In this section, we investigate the tradeoff curve from an optimization perspective.

Based on (\ref{pi-H}) (\ref{PwithPi}) and (\ref{DwithPi}), to achieve the minimum delay given a certain power constraint $P_{th}$, we formulate the following optimization problem
\begin{align}
\min\limits_{\boldsymbol{F},\boldsymbol{\pi}}\quad
& \frac{1}{\alpha A}\boldsymbol{d}^T\boldsymbol{\pi}-1\\
\text{s.t.}\quad
& \boldsymbol{p}_{\boldsymbol{F}}^T\boldsymbol{\pi} \le P_{th}\\
& \boldsymbol{H}_{\boldsymbol{F}}\boldsymbol{\pi}=\boldsymbol{c}\\
& \boldsymbol{\pi} \succeq \boldsymbol{0}\\
& f_{k,m}=0 \qquad \forall k-m<0 \text{ or } k-m>K-A
\end{align}
where $\boldsymbol{\pi} \succeq \boldsymbol{0}$ means $\boldsymbol{\pi}$ is componentwise nonnegative. By varying the value of $P_{th}$, the corresponding optimal delay will be obtained. Hence the optimal delay-power tradeoff curve can be acquired.

Define $x_{k,m}=\pi_k f_{k,m}$. From (\ref{sum=1}), (\ref{PwithPi}) and (\ref{DwithPi}), we can express the average delay and power using $x_{k,m}$ as
\begin{align}
P=\sum_{k=0}^{K} \sum_{m=0}^M P_m x_{k,m}
\end{align}
and
\begin{align}
D=\frac{1}{\alpha A}\sum_{k=0}^{K} k \sum_{m=0}^M x_{k,m}-1.
\end{align}

From global equilibrium equations
\begin{align}
&\sum_{l=\max\{0,k-A\}}^{k-1}\sum_{m=0}^{l+A-k}\alpha \pi_l f_{l,m}\nonumber\\
=&\sum_{r=k}^{\min\{k+M-1,K\}}\sum_{m=r-k+1}^{r-k+A} (1-\alpha) \pi_r f_{r,m}\nonumber\\
&+\sum_{r=k}^{\min\{k+M-1,K\}}\sum_{m=r-k+A+1}^M \pi_r f_{r,m} \qquad k=1,\cdots,K\label{equi_f}
\end{align}
we have
\begin{align}
&\sum_{l=\max\{0,k-A\}}^{k-1}\sum_{m=0}^{l+A-k}\alpha x_{l,m}=\sum_{r=k}^{\min\{k+M-1,K\}}\sum_{m=r-k+A+1}^M x_{r,m}\nonumber\\
&+\sum_{r=k}^{\min\{k+M-1,K\}}\sum_{m=r-k+1}^{r-k+A} (1-\alpha) x_{r,m} \qquad k=1,\cdots,K.\label{equi_x}
\end{align}

The hazard of overflow or underflow can be eliminated by setting $f_{k,m}=0$ if $k-m<0$ or $k-m>K-A$, hence $x_{k,m}=0$ if $k-m<0$ or $k-m>K-A$. Moreover, since $\sum_{k=0}^{K}\pi_k=1$, we have $\sum_{k=0}^{K} \sum_{m=0}^M x_{k,m}=1$. Therefore the original optimization problem can be transformed to
\begin{align}
\min\quad
& \frac{1}{\alpha A}\sum_{k=0}^{K} k \sum_{m=0}^M x_{k,m}-1\\
\text{s.t.}\quad
& \sum_{k=0}^{K} \sum_{m=0}^M P_m x_{k,m} \le P_{th}\\
& \sum_{l=\max\{0,k-A\}}^{k-1}\sum_{m=0}^{l+A-k}\alpha x_{l,m}\nonumber\\
&=\sum_{r=k}^{\min\{k+M-1,K\}}\sum_{m=r-k+A+1}^M x_{r,m}\nonumber\\
&+\sum_{r=k}^{\min\{k+M-1,K\}}\sum_{m=r-k+1}^{r-k+A} (1-\alpha) x_{r,m} \quad k=1,\cdots,K\\
& \sum_{k=0}^{K} \sum_{m=0}^M x_{k,m}=1\\
& x_{k,m}=0 \qquad \forall k-m<0 \text{ or } k-m>K-A\\
& x_{k,m}\ge 0 \qquad \forall 0 \le k-m \le K-A.
\end{align}

It can be observed that this is a linear programming problem. Based on mature algorithms such as simplex method or interior point method, the optimization can be well solved.

\section{Numerical Results}
In this section, we validate our theoretical conclusions by conducting numerical computation and simulation. It will be confirmed that the optimal delay-power tradeoff curve obtained by the Pareto frontier of the feasible delay-power region and the one obtained by linear programming are the same.

\begin{figure}[t]
\centering
\includegraphics[width=1\columnwidth]{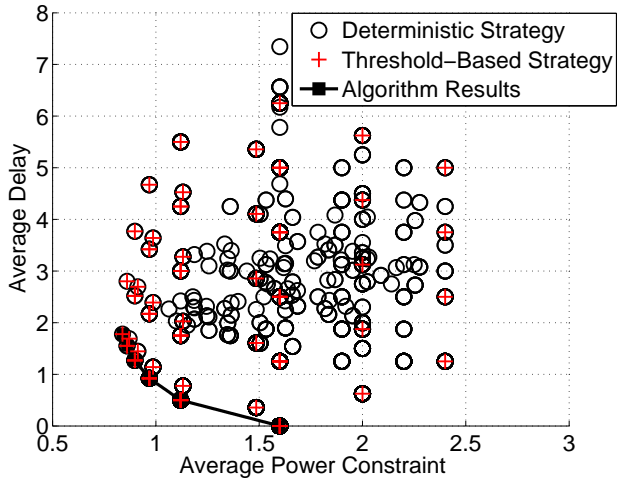}
\caption{Delay-Power Pairs Generated by Deterministic Strategies}
\label{fig_strategy}
\end{figure}

In \figurename \ref{fig_strategy}, we show all the delay-power pairs generated by deterministic strategies. The parameters are $K=7$, $A=2$, $M=3$, $\alpha=0.4$, $P_0=0$, $P_1=1$, $P_2=4$, $P_3=9$. Hence the feasible delay-power region is the convex hull of all the points. Within these delay-power pairs, we mark the ones that are corresponding to threshold-based strategies with '+' markers. As we can see, Algorithm \ref{algo} obtains the Pareto frontier of the feasible delay-power region, which is the optimal delay-power tradeoff curve. The optimal tradeoff curve is piecewise linear, whose vertices are all obtained by threshold-based deterministic strategies.

\begin{figure}[t]
\centering
\includegraphics[width=1\columnwidth]{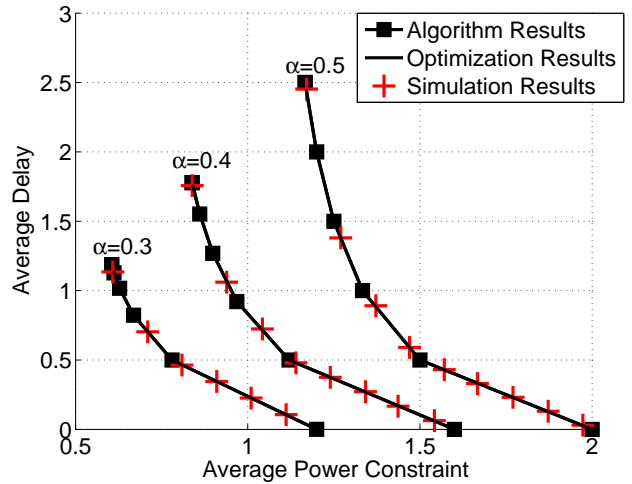}
\caption{Optimal Delay-Power Tradeoff Curves}
\label{fig_optimization}
\end{figure}

In \figurename \ref{fig_optimization} where $K=7$, $A=2$, $M=3$, $P_0=0$, $P_1=1$, $P_2=4$, $P_3=9$, we demonstrate that the optimal delay-power tradeoff curve obtained by linear programming completely overlap the optimal tradeoff curve generated by Algorithm \ref{algo}. The results are further validated by Monte-Carlo simulation. As is expected, the average delay decreases when the average power consumption increases. When the power constraint is no less than $\alpha P_A$, which corresponds to the initial strategy of Algorithm \ref{algo}, the average delay decreases to 0 because any data will be immediately transmitted. With $\alpha$ increasing, the curve gets higher because of the heavier workload.

\section{Conclusion}
In this paper, we investigate the buffer-aware probabilistic scheduling strategies and the corresponding average delay and power consumption. Specifically, random data arrival and adaptive transmission rate are considered. The average delay and power consumption are analysed by Markov reward process. Based on this, we discover that the feasible delay-power region is a polyhedron. The optimal delay-power tradeoff curve, as the Pareto frontier of the polyhedron, is piecewise linear, whose vertices are obtained by deterministic strategies. Furthermore, the optimal delay-power tradeoff is achieved by threshold-based strategies, thus an effective algorithm is proposed to obtain the optimal tradeoff curve. Moreover, seen from another angle, we formulate a linear programming problem to minimize the average delay given a fixed average power constraint. The optimal delay-power tradeoff curves obtained by the algorithm and the optimization problem match perfectly well, which are further validated by Monte-Carlo simulation.


\bibliographystyle{IEEEtran}
\bibliography{IEEEabrv,scheduling}

\end{document}